# Impact of Errors in Operational Spreadsheets

Stephen G. Powell, Barry Lawson, and Kenneth R. Baker
Tuck School of Business
Dartmouth College
Hanover NH 03755 USA

May 22, 2007

**Abstract**
All users of spreadsheets struggle with the problem of errors. Errors are thought to be prevalent in spreadsheets, and in some instances they have cost organizations millions of dollars. In a previous study of 50 operational spreadsheets we found errors in 0.8% to 1.8% of all formula cells, depending on how errors are defined. In the current study we estimate the *quantitative impacts* of errors in 25 operational spreadsheets from five different organizations. We find that many errors have no quantitative impact on the spreadsheet. Those that have an impact often affect unimportant portions of the spreadsheet. The remaining errors do sometimes have substantial impacts on key aspects of the spreadsheet. This paper provides the first fully-documented evidence on the quantitative impact of errors in operational spreadsheets.

## 1. Introduction

Although previous research has suggested that errors are prevalent in spreadsheets (Panko and Halverson, 1996; Panko 2005), we have only limited information on the types of errors that occur, why they occur, and how to avoid them. We recently summarized the research literature on spreadsheet errors (Powell et al., 2006a) and came to the following conclusions:

- There is no generally agreed upon taxonomy of errors in spreadsheets.
- The distinction between actual errors and poor practices is not well-defined.
- All studies that have examined spreadsheets in the field have found errors, although there is no standardization in these studies of
    - the categories of errors
    - the methods used to detect errors
    - the sample of spreadsheets studied.
- It is not possible at present to estimate accurately the prevalence or the impact of errors in spreadsheets.

The research we report here was designed to test for errors in a large sample of spreadsheets in actual use by organizations. We developed an explicit auditing protocol and trained a group of researchers to apply it consistently. Using an explicit protocol is important for two reasons: one, it allows other researchers to replicate and improve on our work; two, it contributes to the development of improved auditing procedures, which is important in its own right. The auditing procedure is described in detail in Powell et al. (2006b).

We begin this paper with a brief review of previous work on spreadsheet errors. Then we describe the design of our study and the sample of spreadsheets we audited. In a previous study we audited 50 spreadsheets from a variety of organizations and summarized the results in terms of *error instances* (the occurrence of a single type of error) and *error cells* (the number of cells affected by a single error instance). The current study focuses on the *quantitative impact* of errors. Each error identified in this study was verified by the developer of the spreadsheet, and the quantitative impact of fixing the error was determined.

## 2. Research design

Our research into spreadsheet errors is focused on *completed, operational* spreadsheets, not errors made in a laboratory setting or errors made during the development of a spreadsheet. In a previous study we used an auditing procedure on a large number of spreadsheets, but we did not have access to the developers. This allowed us to audit a large number of spreadsheets, but it also meant that we could not check our understanding of a model with the developer. In practice, this meant that at times we accepted a suspicious formula as correct because we could not be *sure* that it was incorrect. In the current study we worked with 25 volunteer spreadsheet developers within five different organizations. The developers each completed a survey describing their spreadsheet's purpose and design, as well as any unusual or special formulas or assumptions. Two researchers then independently audited each spreadsheet and pooled their results. Finally, we debriefed the developer on each issue our audit raised, and categorized each issue as an error, a poor practice, or not an error. Actual errors were then corrected and the change in the relevant output cell was recorded as the quantitative measure of the impact of the error.

*2.1 Sample spreadsheets*

We first identified five organizations that were willing to provide volunteers and to have their spreadsheets audited. This included two consulting companies, a large financial services firm, a manufacturing company, and a college. Each organization identified five volunteers, each of whom provided one spreadsheet for auditing. We provided the following specifications to the volunteers for help in choosing a spreadsheet to audit:

- contains 3-10 worksheets
- contains 250-10,000 formulas
- occupies 200-1,000 kb of memory
- developed by the volunteer in the last twelve months
- contains no complex Visual Basic code
- is well understood by the developer
- has no broken external links

Not all the spreadsheets we audited conform to these specifications. In fact, the average number of sheets in our sample was 15.2 (the range was 2 to 135 sheets) and the average size in kilobytes was 1,463 (the range was 45 to 7,450kb). Many of the spreadsheets in our sample were larger on one or more of the dimensions than specified above.

While our sample is not strictly *random*, it is certainly *representative* of the general population of spreadsheets (with the caveats cited above). The sample includes spreadsheets from different types of organizations, spreadsheets created by developers with different degrees of expertise, and spreadsheets that span a broad range from small and simple to large and complex.

*2.2 Auditing protocol*

The auditing protocol we developed for a previous study is a highly-detailed document that specifies the steps to take in auditing a spreadsheet for size, complexity, several types of qualitative features, and errors. (A complete description of the protocol is available at our research project's website, http://mba.tuck.dartmouth.edu/spreadsheet/index.html.)

This protocol evolved over several months of testing. During this time we trained auditors and tested the protocol ourselves and through our assistants on dozens of operational spreadsheets. Our current study used a very similar approach but focused less on gathering data about the spreadsheets and more on finding potential errors. The typical approach was to review the survey provided by the developer, especially the portion describing the various worksheets and their interrelationships. The next step was to run the auditing software *Spreadsheet Professional* (http://www.spreadsheetinnovations.com), which provides summary maps for each sheet and reports on the location of potential errors. We then examined each sheet in turn, first looking at the map for suspicious cells or ranges and scanning the error tests for unusual conditions. Then we inspected the sheet itself, first determining the location of data and formulas and subsequently auditing every unique formula and most copied formulas.

In this study we recorded data only on cells that were potentially errors. For each problematic cell or range we recorded the following information:

- location: cell address or range
- type of error (see below)
- how it was discovered (whether by map analysis, error tests, code inspection, or sensitivity testing)
- description of the possible error

After we had discussed the potential error with the developer, we then recorded how the issue was resolved (Error, Poor Practice, No Error). For Errors, we also recorded the cell used to measure the impact and the absolute and percentage changes in that cell when the error was corrected.

*2.3 Measuring impacts*
Measuring the impact of an error on a spreadsheet is necessarily somewhat subjective. First, some errors occur in a single cell while others occur in many cells. Do we consider each cell separately and measure the impact of correcting it alone, or do we correct all cells with a similar error and measure the overall impact? Second, some error cells are used to calculate many other cells while others impact no other cells. When a cell impacts many other cells it is not always obvious which of the impacted cells to use to measure the effect. (And, of course, different errors can impact different cells.) Third, it is not always clear how to correct an error. For example, if erroneous inputs were used do we replace them with average inputs or extreme inputs? Finally, it is necessary to decide whether to measure errors in absolute or relative terms, and whether to combine all the errors in a given workbook into one overall error.

In this study we chose to measure the effect of each error separately. In most cases we corrected all cells with a given type of error, considering this one error with a single impact. When such an error impacted only the erroneous cells themselves we computed the *maximum* change from the base case and took that as our error estimate. When such an error impacted a single cell we measured the impact of correcting all the error cells on that one cell. We did not attempt to determine a single error estimate for each workbook but measured each error separately. In many cases the only cell impacted by an error was the cell itself. When an error cell had dependencies, we traced these to what we judged to be the most important dependent cell and measured the impact of correcting the error on that cell.

When we began this research, we had expected that most spreadsheets would be set up to calculate a few key outputs. If this were the case, it would be straightforward to determine the impact on these key outputs of an error anywhere in the spreadsheet. We were surprised to discover during our research that many spreadsheets do not have a small number of key output cells. Rather, we found many examples in which hundreds or thousands of results were calculated that themselves had no dependents. This made it more difficult to determine the impact of errors.

In the end we had to use quite a bit of judgment to decide which cell or cells to use to measure the impact of errors.

**3. Error types**
One of the challenges of spreadsheet error research is how to categorize errors. As we pointed out earlier, many different error classifications have been offered (Rajalingham, et al., 2000; Purser and Chadwick, 2006). Most of these suffer from the same flaw: errors that arise from different causes cannot be distinguished by an auditor. For example, when we encounter an error in a formula we can rarely determine whether the error was due to sloppy typing, lack of domain knowledge, lack of Excel knowledge, a subsequent user changing the formula, or an unknown cause. We can, however, easily detect some formulas that give the wrong result. We can also identify many practices that are likely to cause errors as the spreadsheet is used or that simply will make it harder than necessary to use the spreadsheet productively. Other poor practices include limited or nonexistent documentation, duplication of inputs, illogical physical layout, and so on.

After considerable testing we settled on the following six error types that our experience with auditing suggested were well-defined in theory and could be identified with high reliability in practice:

- Logic error - a formula is used incorrectly, leading to an incorrect result
- Reference error - a formula contains one or more incorrect references to other cells
- Hard-coding numbers in a formula - one or more numbers appear in formulas and the practice is sufficiently dangerous
- Copy/Paste error - a formula is wrong due to inaccurate use of copy/paste
- Data input error - an incorrect data input is used
- Omission error - a formula is wrong because one or more of its input cells is blank.

More information on this error taxonomy is available in Powell et al. (2006c).

We should point out that hard coding was identified in our previous study as the most common poor practice. In this study we ignored hard coding in most instances, on the grounds that it was unlikely to represent an outright error. However, there were a few instances in which contradictory inputs were hard-coded and we did cite those as potential errors.

**4. Impact of errors**
Table 1 summarizes our results. In column 1 we have used a two-digit code to label each spreadsheet. For example, spreadsheet 3.4 is the fourth spreadsheet from organization 3. The table gives the following information for each spreadsheet:

- number of issues raised in our audits
- number of errors confirmed in interviews
- number of errors with non-zero quantitative impact
- maximum percentage impact
- maximum absolute impact

Within this sample of 25 spreadsheets we identified a total of 381 issues. After we discussed these issues with the developers we found that nine spreadsheets had no errors; among the remaining 16 spreadsheets we found a total of 117 errors. Of these 177, 47 had zero quantitative impact, leaving 70 errors with non-zero impact.

As we pointed out above, there are two ways to measure the impact of errors: absolute and relative. Absolute impacts are important because they tell us how large errors are in the units of the spreadsheet. However, they cannot be compared easily across workbooks, since a million dollar error may be trivial in one spreadsheet and catastrophic in another. Relative (or percentage) errors more accurately reflect the significance of an error, but they have their shortcomings as well. One problem with relative errors is that percentage changes cannot be determined when the initial value is zero; another is that percentage changes in percentages are not generally as meaningful as percentage changes in dollar amounts. We present our results here in both absolute and relative terms.

Figure 1 shows the distribution of absolute error impacts. The most salient point to draw from this figure is that 47 of the errors we found had zero impact on the spreadsheet. This often came about when a formula had an erroneous reference, but both the erroneous and the correct input cells had the same value. Thus when the error was fixed the results did not change.

Returning to Figure 1, we see that 12 of the errors involved percentages; among these the average absolute change was 22%. Twenty-four of the remaining 58 errors involved absolute errors less than $10,000. However, some errors were huge: the largest single absolute error we found was over $100 million!

Figure 2 shows the distribution of percentage error impacts. (The NA category includes four errors in cells with an initial value of zero, for which a percentage change is not defined.) Of course 47 of the 117 errors had zero impact, regardless of how measured. Forty-seven of the 66 remaining errors were less than 10% of the initial value. As with absolute errors, there are some very large errors: four, in fact, were over 100%.

Our evidence suggests that spreadsheet practice is very different among the five organizations we studied. In the five spreadsheets from Organization 5 we could identify only five issues to discuss with the developers and no errors were identified among those five issues. Organization 5 is a small consulting company with highly educated employees and a culture that demands excellence. The spreadsheets we audited from this firm were works of art: thoughtfully designed, well documented, easy to understand, and error free.

Organization 4 had two spreadsheets with no errors and two with 22 and 44 errors, respectively. The quality of the spreadsheet practice in this organization clearly depends on just where one looks. In this case we found both the best of practice and the worst of practice in offices just a few miles apart.

In Organization 3, which is another consulting company, all the spreadsheets we audited had errors but in three cases no error had a measurable impact on the results. Even in the remaining two spreadsheets the errors were few in number and fairly small in terms of impact.

Organizations 1 and 2 are both very large. One is a financial firm and the other is a manufacturing firm. Some of the spreadsheets we audited from these companies were astonishingly large and complex. Perhaps for this reason, only two of the ten we audited were error-free (four had no errors with impact). The quality of spreadsheet practice in both of these companies was inconsistent, with inadequate attention paid to spreadsheet design, simplicity, ease of use, documentation, and consistency.

We can summarize our findings as follows:

- Some organizations have spreadsheets that are essentially error-free.
- Within a single organization, spreadsheet practice can range from excellent to poor.
- Some organizations use spreadsheets that are rife with errors and some of these errors are of substantial magnitude.
- Many errors have zero impact, or impact unimportant calculations.
- There is little correlation between the importance of the application or the risk involved and the quality of the spreadsheet.
- Few spreadsheets contain errors that, in the eyes of their developers, would destroy their usefulness.

**5. Qualitative observations**
Many writers have observed that the spreadsheet, for all its attractiveness to end-users, is in some ways a dangerous modeling platform. Not only are the logic of the model and the numbers commingled, but the physical layout permits unstructured designs. It is not surprising that amateur programmers, who lack structured design methods, make errors when using such free-form software.

Our research has shown that errors come in more varieties than perhaps even the most extensive taxonomy can encompass. Because the spreadsheet platform is so unstructured, and because end-users generally follow unique designs, errors and poor practices can arise in thousands of different guises. Error researchers inevitably must use their judgment in deciding what is an error and what is not. Thus we should be skeptical of claims about the frequency and impact of errors based on rough averages and casual research.

Another general observation that our research supports is that spreadsheet auditing is more difficult and limited than might have been anticipated. We knew in advance that we would not be able to identify errors in problem formulation or in the use of the model by auditing the spreadsheet itself, although these types of errors may be the most consequential. But even within the narrower domain of our audits we encountered limitations. First, the data used in most spreadsheets is undocumented and there is no practical way to check it. Even the original developer would have difficulty checking the data. Second, many spreadsheets are so chaotically designed that auditing (especially of a few formulas) is extremely difficult or impossible. Finally, we have found that many spreadsheets do not have just a few key outputs but are used to calculate hundreds or thousands of results. This makes it difficult to unambiguously measure the impact of a particular error.

One important generalization our work supports is that many errors are benign: they either have no impact on the results, the quantitative impact is very small, or the effect is on a vestigial portion of the spreadsheet that is no longer of interest. One can conjecture that this is the result of a sensible attitude toward errors on the part of spreadsheet developers. Perhaps developers look out for errors that impact the key outputs, and in general are good at correcting them. However, they pay less attention to inconsequential errors and therefore more of these survive to be observed. And, as we know from our interviews, many developers do not clean up their spreadsheets before they move on to other tasks.

One factor that might explain the substantial differences within and among companies in the quality of their spreadsheets is the degree of risk involved. We might hypothesize that companies devote their best resources to high-risk spreadsheets, and fewer resources to low-risk ones. (There is some evidence in a user survey we conducted that would support this conjecture. See Lawson et al. (2006).) We did not measure this feature of the spreadsheets we audited (this would certainly be difficult to do), but our impression is that no such correlation existed within our sample. For example, one of the best spreadsheets we audited was designed to help with daily staffing of nurses and doctors to a medical practice. The spreadsheet was elegantly engineered, error-free, and easy to use. But little was at risk: an error in staffing at worst would assign the same person to adjacent shifts or to too many consecutive days. Errors of this type would almost certainly be caught, and their impact would be negligible. None-the-less, the spreadsheet was nearly perfect. By contrast, we also audited spreadsheets in use in a major financial firm for calculating tax liabilities (measured in the billions) to various state and national entities. These spreadsheets were astonishingly complex, difficult to understand, difficult to work with, and error-prone. So factors other than risk appear to explain spreadsheet quality.

Another observation that helps to understand our results is that many of the developers we worked with were not especially surprised or devastated when we pointed out potential errors. Sometimes the reaction was that they knew the formula "wasn't quite right" but they saw that it gave the right *answer* and thus was acceptable. Sometimes the reaction was that the result was "close enough," or that the result in question was no longer used, or not important. So developers seem to have a sense of what level of accuracy is appropriate for a given spreadsheet. (It is another question entirely as to whether their perceptions are correct, and the spreadsheets are actually as accurate as they need to be.)

An experienced auditor can rather quickly detect a spreadsheet that is likely to have errors. The major symptoms we observed of poor spreadsheet practice are the following:

- Chaotic design
- Embedded numbers
- Special cases
- Non-repeating structures
- Complex formulas.

Chaotic design refers to a poorly structured physical layout of the formulas and data. Numbers embedded in formulas, while not necessarily direct causes of errors, are strongly correlated with the presence of other problems. Special cases refers to designs in which similar results are calculated in slightly different ways, which requires great care in building and checking formulas. Non-repeating structures includes designs in which the formulas in a row or column change structure repeatedly, precluding the use of copying and pasting. In the hands of experts, complex formulas can be used to great effect. But in the hands of novices the same formulas can be error-prone.

Why is spreadsheet practice sometimes so poor? We cannot know for sure, but we did gather some anecdotal evidence during our interviews. When asked what kept them from building better spreadsheets, our developers typically cited one or more of the following reasons:

- Time pressure
- Organic design
- Changing specs
- Lack of testing
- Lack of relevant knowledge and skills

Time pressure was the most often cited reason. Many spreadsheets are built under great time pressure, which precludes use of some of the most effective methods for avoiding errors. Managers of spreadsheet developers should be aware of the effects of putting their employees under excessive time pressure. Another commonly cited factor was organic design: either the spreadsheet design was inherited from a predecessor spreadsheet, or it grew organically during the project without ever consciously being designed. Another complaint was changing specifications: if the designer had only known from the start what the spreadsheet was going to used for, he or she could have designed it more appropriately. We also observed that very few of our developers used any formal approach to testing their spreadsheets; in fact, most of them did no testing as such. Finally, in some cases we could observe directly that the cause of an error was lack of relevant knowledge, either of the problem domain or of spreadsheet tools. It was remarkable, however, how rare this cause appeared to be. Most developers could see quickly that a particular formula was an error, once we had pointed it out to them. Only very rarely did we have to explain to them why it was an error, or how to fix it.

Finally, we offer a comment on the importance to auditing, and to good spreadsheet usage in general, of good design. Our work makes us extremely conscious that a well-designed spreadsheet is *simple*, *consistent*, and *general*. Simplicity means a logical use of worksheets and a logical and intuitive layout of each individual sheet. Simplicity makes building and auditing easier. Consistency means, for example, that a single formula can be written and then copied down an entire row or across an entire column. Such a formula can easily be checked. Rows or columns in which the formulas change structure constantly often hide errors. Generality means that the spreadsheet is built to handle all of the likely combinations of inputs that users will want to use. The opposite is a workbook in which individual cases are calculated separately, which makes it difficult to keep inputs consistent across cases.

**6. Summary and Future Research**
We have audited 25 spreadsheets from five organizations. We identified cells or ranges that appeared to be problematic and discussed each one with the developer of the spreadsheet. For each issue that was classified as an actual error, we then identified the cell or cells affected and measured the absolute and percentage impact of correcting the error. Several conclusions emerge from this research:

- The quality of spreadsheet practices differs substantially among and within organizations.
- Some individuals and organizations are capable of developing essentially error-free spreadsheets.
- Many spreadsheets are built in ways that violate good design practices.
- Operational spreadsheets are highly complex and often poorly structured.
- Poor practices (such as hard-coding numbers in formulas) abound, but quantitative errors are relatively rare.
- The quantitative impact of many errors is negligible or zero, or occurs in unimportant cells.
- Devastating errors are rare.

None of these conclusions should be taken as proven by the current research. Rather, they are suggestive hypotheses that should be refined through further research.

| Organization-Workbook | # Issues | # Errors | Errors with Non-zero Impact | Maximum Percentage Impact | Maximum Absolute Impact |
|---|---|---|---|---|---|
| 1.1 | 7 | 3 | 0 | NA | NA |
| 1.2 | 50 | 6 | 5 | 28.8% | $32,105,400 |
| 1.3 | 18 | 7 | 3 | 137.5% | $110,543,305 |
| 1.4 | 4 | 1 | 0 | NA | NA |
| 1.5 | 0 | 0 | 0 | NA | NA |
| 2.1 | 19 | 6 | 5 | 3.6% | $13,909,000 |
| 2.2 | 27 | 11 | 7 | 16.0% | $74,000,000 |
| 2.3 | 6 | 0 | 0 | NA | NA |
| 2.4 | 30 | 4 | 3 | 416.5% | $10,650,000 |
| 2.5 | 40 | 2 | 2 | NA | 8.90% |
| 3.1 | 19 | 2 | 2 | 5.3% | $238,720 |
| 3.2 | 1 | 1 | 0 | NA | NA |
| 3.3 | 11 | 2 | 2 | 15.6% | $4,930,000 |
| 3.4 | 6 | 1 | 0 | NA | NA |
| 3.5 | 23 | 1 | 0 | NA | NA |
| 4.1 | 27 | 22 | 12 | 116.7% | $13,355,445 |
| 4.2 | 8 | 4 | 2 | 141.8% | $272,000 |
| 4.3 | 0 | 0 | 0 | NA | NA |
| 4.4 | 1 | 0 | 0 | NA | NA |
| 4.5 | 79 | 44 | 27 | 39.1% | $216,806 |
| 5.1 | 2 | 0 | 0 | NA | NA |
| 5.2 | 2 | 0 | 0 | NA | NA |
| 5.3 | 0 | 0 | 0 | NA | NA |
| 5.4 | 0 | 0 | 0 | NA | NA |
| 5.5 | 1 | 0 | 0 | NA | NA |
| Totals | 381 | 117 | 70 | | |

**Table 1**
**Summary of Audit Results**

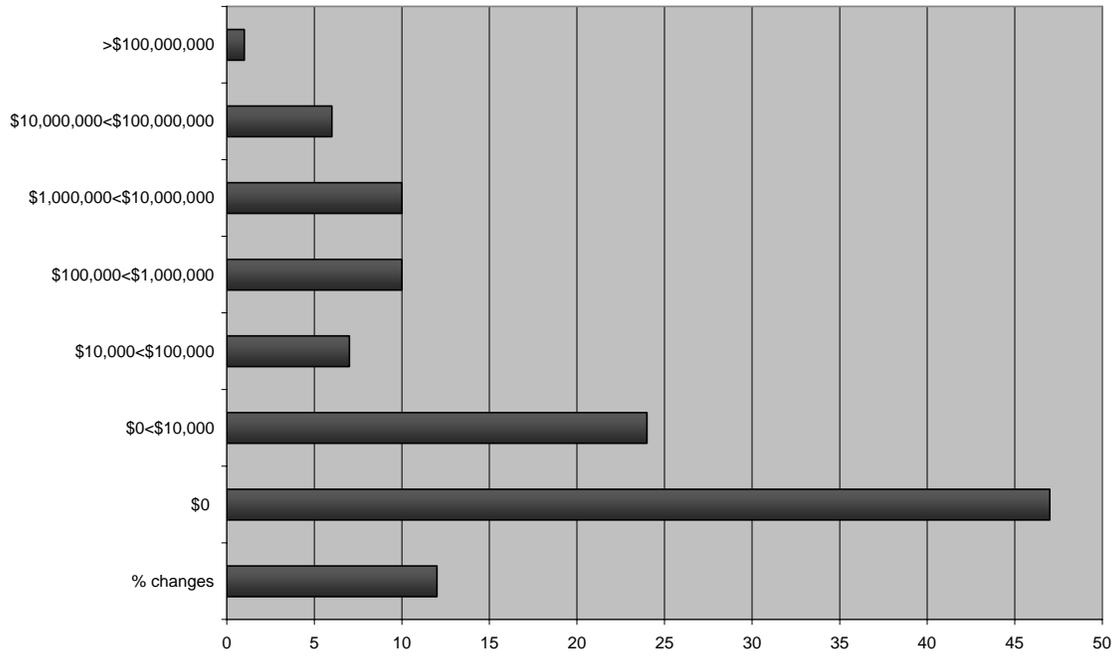

Figure 1. Distribution of Errors, by Absolute Impact
(n = 117)

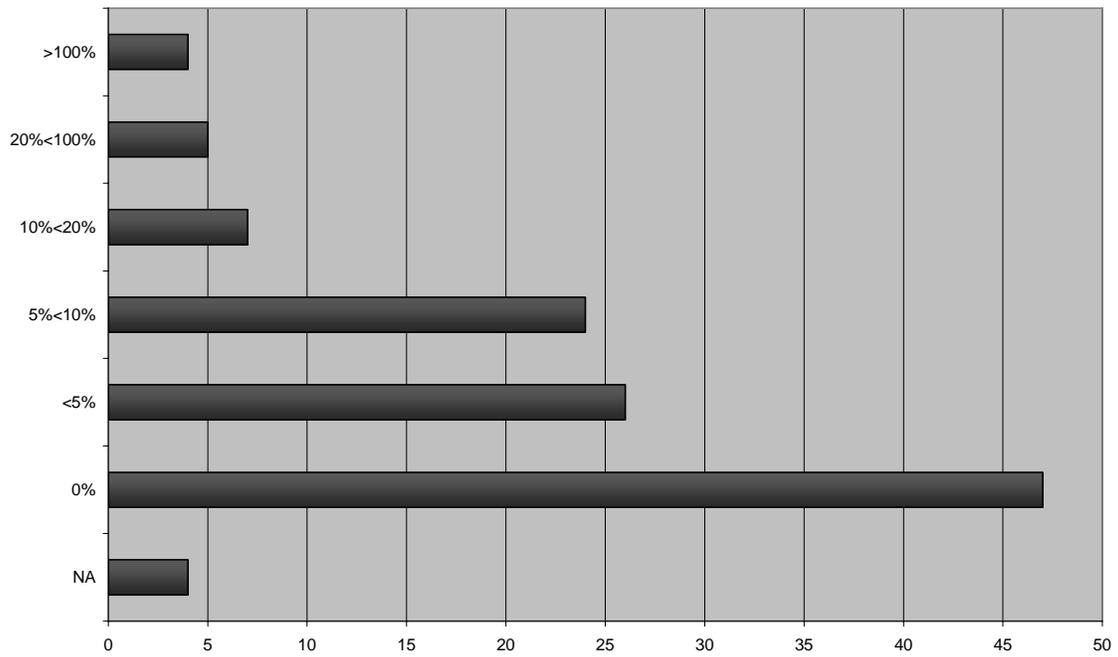

Figure 2. Distribution of Errors, by Percentage Impact
(n = 117)

Blank Page